# A Simple Scheme for Distributed Passive Load Balancing in Mobile Ad-hoc Networks

**Benjamin Sliwa, Robert Falkenberg and Christian Wietfeld**
Communication Networks Institute
TU Dortmund University
44227 Dortmund, Germany
e-mail: {Benjamin.Sliwa, Robert.Falkenberg, Christian.Wietfeld}@tu-dortmund.de

*Abstract*—Efficient routing is one of the key challenges for next generation vehicular networks for providing fast and reliable communication in a smart city context. Various routing protocols have been proposed for determining optimal routing paths in highly dynamic topologies. However, it is the dilemma of those kinds of networks that good paths are used intensively, resulting in congestion and path quality degradation. In this paper, we adopt ideas from multipath routing and propose a simple decentral scheme for Mobile Ad-hoc Network (MANET) routing, which performs passive load balancing without requiring additional communication effort. It can easily be applied to existing routing protocols to achieve load balancing without changing the routing process itself. In comprehensive simulation studies, we apply the proposed load balancing technique to multiple example protocols and evaluate its effects on the network performance. The results show that all considered protocols can achieve significantly higher reliability and improved Packet Delivery Ratio (PDR) values by applying the proposed load balancing scheme.

## I. INTRODUCTION

Future smart cities will lead to a significant growth in the amount of Vehicle-to-Infrastructure (V2I) and Vehicle-to-Vehicle (V2V) communication [1]. As the network topology is frequently changing due to the mobility behavior of the vehicles, MANET routing protocols are a promising candidate for finding reliable communication paths. However, most current mesh routing protocols focus on the determination of a single optimal communication path between sender and destination. High data traffic demands lead to an intense usage of those paths, decreasing the Quality of Service (QoS)-capabilities of the applications due to packet collisions and congestion. To overcome these issues, various *load balancing* strategies have been proposed, which make use of additional coordination and communication in order to signal congested nodes and lower the transmission rates if needed. As these approaches highly raise the protocol complexity, they are rarely applied to real-world mesh routing protocols. In recent work [2], we proposed the novel MANET routing protocol *B.A.T.Mobile*, which uses mobility prediction for optimized forwarding decisions and proactive avoidance of path losses. Every node manages the path quality information of all possible forwarders for a given destination in a score-based routing table. In this paper, we exploit this property to achieve load balancing using a simple packet distribution scheme. The general idea is that spreading data traffic over multiple suboptimal routing paths will lead to a better overall network performance than using a single optimal path. Fig. 1 illustrates the problem statement and shows an example use case for load balancing in the context of V2V communication. However, the proposed method can also be applied in other types of MANETs such as Unmanned Aerial Vehicle (UAV) swarming applications. Our simulation setup has been published as an Open Source framework [3] and is based on the discrete network simulator Objective Modular Network Testbed in C++ (OMNeT++) [4] and its INETMANET framework. The remainder of this paper is structured as follows: after discussing the related work, we present the system model of our solution approach and describe the key components of the suggested load balancing scheme. In the next section, we describe our simulation environment and the reference scenario for the performance evaluation. Finally, detailed results of comprehensive simulation studies are presented and discussed.

## II. RELATED WORK

Load balancing and congestion avoidance have been topics of intense scientific interest for a long time. In the context of wireless mesh networks, there is an overlap to the multipath routing field [5]. Instead of only using a single path for data transmission, protocols exploit the mesh topology to utilize multiple paths at once. This allows intelligent packet distribution strategies or is used as a mechanism of redundancy. A positive side-effect is the inbuilt self-healing property, which enables fast communication recovery after node failures. The downside of maintaining multiple paths in parallel is the increase in communication overhead and protocol complexity, especially for reactive protocols. In [6], the authors present Multipath OLSR (MP-OLSR) as an extension of the well-known proactive routing protocol Optimized Link State Routing (OLSR). The extended version of the protocol provides better QoS properties than regular OLSR. A detailed survey about using multipath knowledge for congestion control is given in [7]. The authors of [8] evaluate a novel geographical load-aware routing protocol in a Vehicular Ad-hoc Network (VANET) context. An experimental evaluation of another traffic-aware protocol for Wireless Mesh Networks (WMNs) is given in [9]. The authors of [10] propose a reactive congestion adaptive routing protocol that uses load estimation to determine an optimal contention window size. In [11], a cooperative load balancing mechanism for cluster-based MANETs is discussed. The state of the art approaches indicate the scientific interest and the relevance of the topic. However, the increase in complexity and communication overhead is a severe disadvantage that is also mirrored by the fact that most current real-world mesh protocols do not implement load balancing mechanisms. Furthermore, most existing load balancing attempts are strongly tied to a specific routing protocol and are not intended for being used in combination with others.

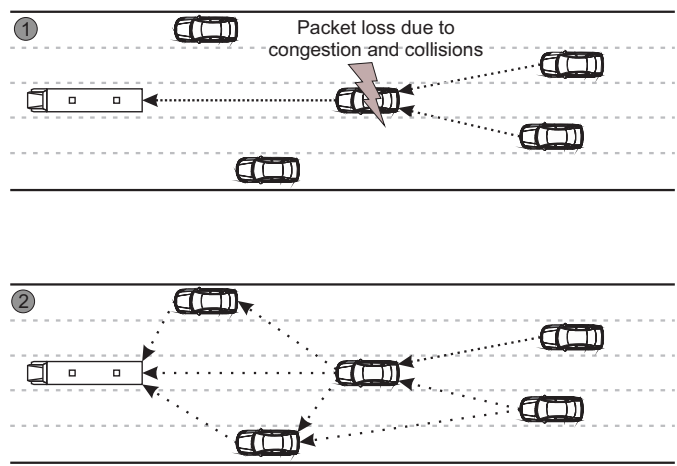

Fig. 1. Example VANET application scenario. In (1), all nodes only use the best routing path which leads packet loss due to congestion and collisions. The network is relived in (2) through message dissemination over all suitable paths.

## III. PASSIVE DECENTRAL LOAD BALANCING

In contrast to existing approaches, we consider only the direct neighbors of a node and the end-to-end transmission path is not taken into account. The packet forwarding decision of every node is only based on local knowledge and statistical assumptions. Fig. 2 shows the system model of our solution approach as an extension to the architecture of the *netfilter* Linux kernel module, where it realizes a *postrouting hook* in order to change the forwarder selection after the routing table has been accessed by the operating system. Due to its simplicity, the load balancing mechanism can easily be applied to existing routing protocols. In the following subsections, we will give a detailed description of the functionality of the key models.

### A. Using Neighbor Rankings for Load Balancing

In order to enable load balancing, nodes require a data structure for managing routing information for all of their one-hop neighbors per destination. Regular routing tables cannot satisfy this requirement as they only maintain single gateway-destination pairs. One possibility is the usage of *score-based neighbor rankings*, which are used by Better Approach To Mobile Adhoc Networking (B.A.T.M.A.N.) and B.A.T.Mobile. With this approach, the suitability of all one-hop neighbors for being used as a packet forwarder towards a defined destination can be compared by a numeric score. This type of data structure is not tied to a specific routing protocol and can be used with any kind of metric. The neighbor rankings act as an extended type of routing table and provide the basis for the proposed load balancing mechanism. The idea is to spread the data traffic over multiple paths that have a similar path quality. However, there is a trade-off between path quality and load balancing. If the optimal path is used intensively, the risk for congestion is high and will cause QoS issues. If paths of too low quality are used, the probability of packet loss is high. We call this trade-off the *likelihood optimization problem* and give a formal description of it in the next subsection.

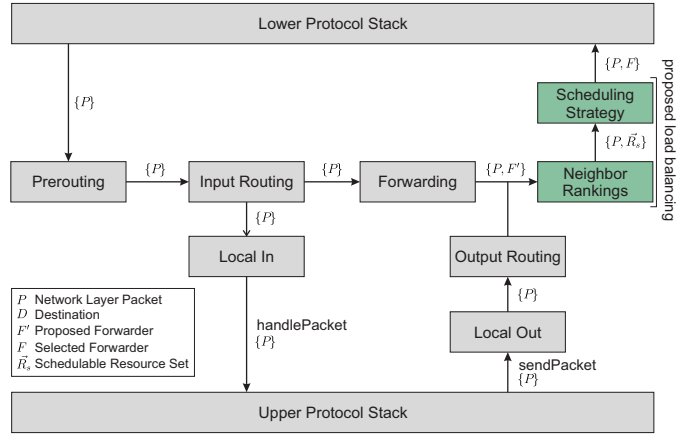

Fig. 2. Architecture model for the netfilter module of the Linux kernel. The components of the load balancing mechanism form a postrouting hook which changes the proposed packet forwarder of the routing table access phase to a load balanced forwarder decision.

### B. Formal Description of the Likelihood Optimization Problem

As an analogy to resource scheduling, we call the one-hop neighbors of a node $s$ the *resource set* $\vec{R}$ of $s$ for a given destination $d$. The task of the load balancing mechanism is to find all one-hop neighbor nodes that are *good enough* for being used as forwarders. Those are a subset of $\vec{R}$ and form the *schedulable resource set* $\vec{R}_s$ of $s$ for $d$. In order to make a decision about the suitability of a one-hop neighbor node $r$ for being used as a forwarder for $d$, $s$ needs a measurement for the path quality of the best path from $r$ to $d$. This *path quality metric* $\Phi$ is provided by the routing protocol itself. Sec. IV-A gives an overview about the metrics of the protocols that are considered in this paper. Using the neighbor rankings, the path quality values of all potential forwarders can be set into relation to the current optimal path with the value $\Phi_{max}$. With the introduction of a *path quality likelihood factor* $\lambda$, $\vec{R}_s$ can now be formally determined using Eq. 1.

$$\vec{R}_s = \left\{ r \in \vec{R} | \Phi(r) > \lambda \cdot \Phi_{max} \right\} \quad (1)$$

The optimal value for $\lambda$ is depending on the routing protocol and its metric and has to be determined through experiments or simulative evaluation. Fig. 3 shows an example network and illustrates the neighbor ranking of node $D$ for the destination $E$. All direct neighbors of $D$ have an entry and an assigned score which serves as an indicator for the path quality to $E$.

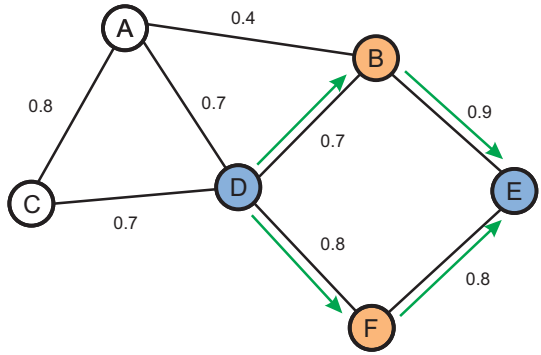

Fig. 3. Example Neighbor Ranking for a path from D to E. The link scores are calculated with the path score metric of B.A.T.Mobile. B and F form the schedulable resource set $\vec{R}_s$ of D for E as they have similar scores for the destination. The nodes obtain the end-to-end scores to other nodes from flooded routing messages.

### C. Message Dissemination

Once $\vec{R}_s$ is obtained, a *scheduling strategy* is applied for distributing the data packets over the schedulable resources. This process is performed by the sender node and all intermediate forwarders. In this paper, we use the simple Round Robin (RR) mechanism for this task that distributes the packets equally over all suitable forwarders. Further extensions could make use of more advanced scheduling strategies and use the knowledge about the path quality for corresponding distribution patterns.

## IV. SIMULATION-BASED SYSTEM MODEL

In this section, we present our simulation-based system model used for the performance evaluation. It consists of the description of the considered routing protocols, the data traffic model and the actual simulation setup.

### A. Routing protocols

Three protocols that make use of different metrics are used for the performance evaluation of the proposed load balancing mechanism. B.A.T.M.A.N. is a bio-inspired routing protocol using the Transmission Quality (TQ)-metric that makes use of penalties for additional hops and serves as an indicator for the transmission success probability. As an extension to B.A.T.M.A.N., we have proposed B.A.T.Mobile in earlier work. Its *PathScore*-metric uses knowledge from the mobility control layer in order to predict future node positions and estimate the distance development between nodes as an indicator for the future link quality. OLSR is a well-established protocol for MANETs. For this paper, we have extended it with a path planning metric based on the geographical distance between sender and destination. This approach provides a finer grained indicator for the path quality than the regular hop count distance and makes it better suited for load balancing. The extended protocol is called Geo-based OLSR (G-OLSR).

### B. Traffic model

The reference scenario is defined by a number of autonomous vehicles moving randomly on a defined playground. This approach causes many changes in the network topology and acts as a stress test for the protocols. One vehicle is randomly selected to continuously stream User Datagram Protocol (UDP) video data to another randomly chosen vehicle.

### C. Routing simulation with OMNeT++/INETMANET

We use the discrete event-based simulation environment OMNeT++ and its INETMANET framework for the evaluation of the proposed load balancing scheme. The scheduling strategy and the neighbor rankings have been implemented as OMNeT++-modules. The load balancing mechanism is triggered by the postrouting hook of the Internet Protocol version 4 (IPv4) implementation. The simulation parameters for the reference scenario are defined in Tab. I. Deviations from the default assignment are explicitly marked when they are required. Tab. II shows the parameters for the routing protocols.

TABLE I
SIMULATION PARAMETERS FOR THE REFERENCE SCENARIO IN OMNET++/INETMANET

| Simulation parameter | Value |
|---|---|
| Mission area | 500m x 500m x 10m |
| Number of agents | [5..25] |
| Mobility model | Controlled Waypoint |
| Velocity $v$ | 50 km/h |
| Channel model | Friis $\gamma = 2.75$ |
| Videostream bitrate | 2 Mbit/s |
| MAC layer | IEEE802.11g |
| Transport layer protocol | UDP |
| MTU | 1460 Byte |
| Transmission power | 100 mW |
| Carrier frequency | 2.4 GHz |
| Receiver sensitivity | -83 dBm |
| Simulation time per run | 600 s |
| Number of simulation runs | 25 |
| Path-quality-likelihood $\lambda$ | 0.9 |

TABLE II
ROUTING PROTOCOL PARAMETERS

| Protocol | Parameter | Value |
|---|---|---|
| B.A.T.M.A.N. | Originator Message (OGM)-interval | 500 ms |
| G-OLSR | HELLO-interval | 500 ms |
| | Topology Control (TC)-interval | 1000 ms |
| B.A.T.Mobile | OGM-intervall | 500 ms |
| | Neighbor Score Buffer size | 8 |
| | Mobility update interval $\Delta t_u$ | 250 ms |
| | Extrapolation data size $N_e$ | 5 |
| | Prediction width $N_p$ | 15 |
| | Grade of relative mobility $\alpha$ | 7 |
| | Maximum path trend $p_{trend,max}$ | 0.1 |

## V. RESULTS OF THE PERFORMANCE EVALUATION

In this section, we present the results achieved with our simulation setup. We consider the PDR of the video stream as our main key performance indicator. Graph plots show the 0.95 confidence interval of the total number of simulation runs.

## A. Performance of the Load Balancing Scheme with different Routing Protocols

In order to identify general characteristics of the effects of the proposed load balancing scheme on the data transmission reliability, we consider the time behavior of regular and load balanced B.A.T.M.A.N. in an example run in Fig. 4 and compare their current PDR values.

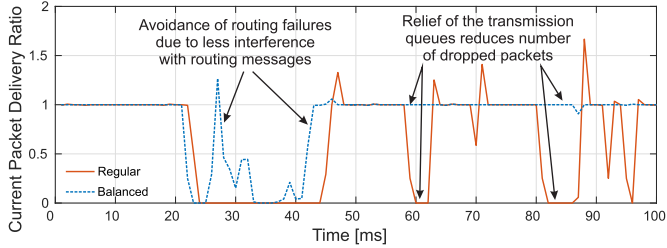

Fig. 4. Example temporal behavior comparison of B.A.T.M.A.N. with and without the load balancing mechanism.

The plain protocol version shows multiple occurrences of current PDR values greater than one, which are related to packet queuing and transmission delays after PDR drops. For the balanced version the graph is much more constant as the transmission queues are relieved through the load distribution. Lost routing packets cause inconsistencies of the routing tables and lead to serious drops of the current PDR in the worst case. Using the proposed distribution scheme for data packets, collisions with routing messages are less probable and route failures can be reduced. A statistical comparison for the con-sidered routing protocols as a plain and a load balanced version is given in Fig. 5. All of them show an increased mean PDR and an improved variance. For G-OLSR and B.A.T.M.A.N. the efficiency gain is most significant as their regular PDR is relatively low and offers a lot of space for improvements. For B.A.T.Mobile the performance improvement is less dramatic.

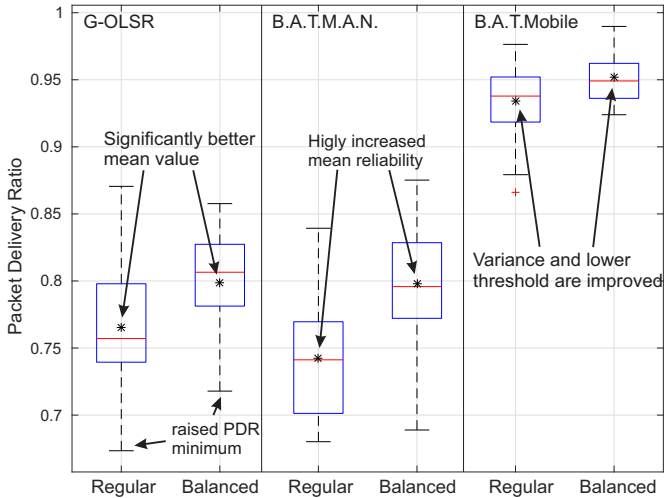

Fig. 5. Comparison of the load balancing effects for different routing protocols.

Using the mobility-aware approach, route changes are triggered proactively with respect to the predicted movement. Due to the high frequency of route changes a simple load balancing is already immanent in the behavior of the protocol. Fig. 6 shows a scalability-analysis. The highest performance gain is achieved for medium-sized groups, which makes it well fitting for applications like platooning and cooperative UAV tasks. As the playground size is not varied for this evaluation the node density is increased with a higher number of vehicles.

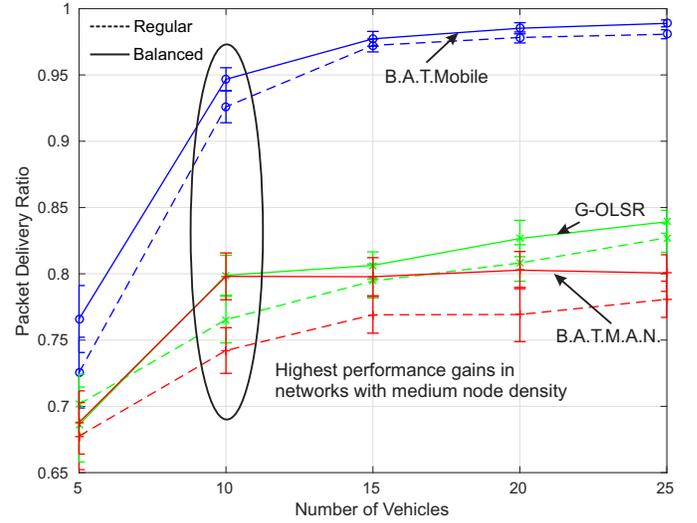

Fig. 6. Load balancing gains depending on the number of vehicles.

## B. Parameterization

The choice of the path quality likelihood factor $\lambda$ is crucial for the performance of the load balancing mechanism. Therefore, we evaluate the effect of different $\lambda$-values for all considered routing protocols in Fig. 7. For $\lambda = 0$ a pure RR is performed, $\lambda = 1$ uses load distribution only if path quality scores are exactly equal and $\lambda > 1$ does not perform any kind of scheduling and is equal to the behavior of the regular protocol. B.A.T.M.A.N. achieves the highest performance gain for medium traffics loads (1-3 parallel streams). With higher amounts of streams the efficiency of the message distribution is reduced as the probability of interfering with another stream is increased. A similar behavior can be seen for G-OLSR but the dependency to the number of streams is higher. As the protocol uses a path planning approach it is highly depending on receiving recent link state information from all network participants. Collision- and congestion-related packet loss therefore decreases the reliability significantly. B.A.T.Mobile achieves the highest absolute PDR but can only benefit slightly from load balancing. The protocol uses the PathScore metric that calculates the total score by multiplication of the scores of the intermediate links. Because of the multiplicative approach, the impact of single links on the total path quality value is high and the metric provides a distinct measurement. Therefore, the scores of different paths will likely differ a lot, which lowers the dependency to an optimal choice of $\lambda$. Although the effects of load balancing are different for the considered protocols, the

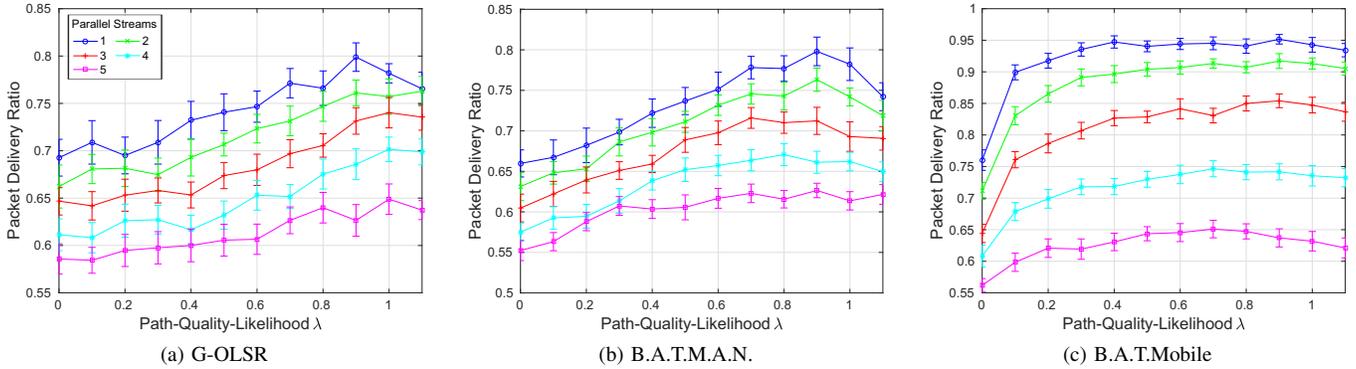

Fig. 7. Impact of the path-quality-likelihood $\lambda$ for multiple MANET routing protocols.

optimal likelihood factor $\lambda$ is similar for all of them. Three different areas can be identified:

- For $\lambda < 0.9$ the PDR is low because low quality links are used by the scheduling strategy.
- For $0.9 \leq \lambda \leq 1.0$ the trade-off between link-quality and load is balanced, which leads to high PDR values.
- For $\lambda > 1.0$ the protocol cannot benefit from load balancing gains as only the best paths are used.

The results show that the path quality estimation has the highest impact on the protocol performance but improvements can be achieved through utilization of decent suboptimal paths. For efficient load balancing, the likelihood of the quality of different paths needs to be very high ($\lambda = 0.9$).

## VI. CONCLUSION

In this paper, we presented a simple passive decentral load balancing approach for MANET routing protocols. In contrast to existing approaches, the nodes consider only local knowledge and no additional communication or coordination is required. The proposed scheme can easily be applied to increase the reliability of existing routing protocols. Our simulative evaluation showed all considered protocols were able to achieve significant PDR gains by integrating the proposed load balancing approach. By distributing the packets over multiple suitable links, packet collisions are less probable and the reliability is increased. The probability for losses of routing packets is reduced, which leads to a higher consistency of the routing tables and avoids occurrences of drastic PDR drops. Additionally, the transmission queues of the forwarding nodes are relieved and queuing-related packet drops occur less often. The highest performance gain can be achieved in medium-dense networks, which are typical for V2V scenarios and cooperative UAV tasks. In future work, we want to analyze the effects of load balancing with further routing metrics and integrate scheduling strategies from cellular networks for better utilization of knowledge about the quality of individual links. Another promising topic is the integration of load information into the calculation of link scores as an adaptive component.

ACKNOWLEDGMENT

Part of the work on this paper has been supported by Deutsche Forschungsgemeinschaft (DFG) within the Collaborative Research Center SFB 876 "Providing Information by Resource-Constrained Analysis", project B4.